©2021.  V.P. Koshcheev[1], Yu.N. Shtanov[2]

# NOISE-INDUCED SELF-OSCILLATION (FLUTTER) SUPPRESSION IN THE KELDYSH MODEL

An equation for the evolution of the energy of a dynamical system (Keldysh model with one degree of freedom), which contains a source of white noise, is constructed. It is shown that self-oscillations (flutter) are suppressed if the intensity of white noise exceeds a critical value.

## 1. INTRODUCTION

More than seventy years ago, nonlinear differential equations were constructed (the Keldysh model with one and two degrees of freedom [1]), the self-oscillating solutions of which made it possible to explain the nature of flutter. The Keldysh model [1] continues to attract the attention of researchers [2]. The question of the influence of noise on the Keldysh model remained open all these years. A similar problem of the effect of noise on the van der Pol oscillator was solved for the first time in [3], to which attention was drawn in [4]. It was shown in [3] that if the noise intensity exceeds a certain critical value, then the self-oscillations disappear. A similar result was obtained in [5] using the stochastic equation for the evolution of the energy of a dynamical system with one degree of freedom [6]. In this paper, the stochastic equation for the evolution of the energy of a dynamical system with one degree of freedom [6] is applied to the Keldysh model with one degree of freedom, which includes an additive source of white noise.

## 2. SOLUTION OF A NONLINEAR DIFFERENTIAL EQUATION IN THE KELDYSH MODEL

The nonlinear differential equation (see, for example, the Keldysh model with one degree of freedom [2]) has the form

$$J\ddot{x} + kx = -\mu\dot{x} - (\Phi + \kappa\dot{x}^2)\operatorname{sign}\dot{x}, \qquad (1)$$

where constants $J = k = \kappa = 1, \mu < 0, \Phi > 0$ are defined in [2].

If the energy of an unperturbed dynamical system has the form

$$E = \frac{J\dot{x}^2}{2} + \frac{kx^2}{2}, \qquad (2)$$

[1] Moscow Aviation Institute (National Research University), Strela Branch, Moscow oblast, Zhukovskii, Russia.
E-mail: koshcheev1@yandex.ru;
[2] Industrial University of Tyumen, Surgut Branch, Surgut, Russia. E-mail: yuran1987@mail.ru

then the trajectory turning points have the form

$$x_{1,2} = \pm\sqrt{\frac{2E}{k}}. \tag{3}$$

The stochastic equation for the evolution of the energy of a dynamical system with one degree of freedom [6] has the form

$$\frac{dE}{dt} = \frac{2}{T}\int_{x_1}^{x_2}(\bar{f} + \delta f)dx, \tag{4}$$

where $T = 2\pi\sqrt{\frac{J}{k}}$ – oscillation period of the unperturbed system; $\bar{f} = -\mu\dot{x} - (\Phi + \kappa\dot{x}^2)\operatorname{sign}\dot{x}$; $\delta f = \sqrt{2D}\xi(t)$; $D$ – the intensity of the random source noise; $\xi = \xi(t)$ – random variable with unit variance.

We find the speed of the dynamic variable using (2)

$$\dot{x} = \pm\sqrt{\frac{2E - kx^2}{J}}. \tag{5}$$

Using (5), we rewrite equation (4) in the form

$$\frac{dE}{dt} = \mp(a_0 E^{\frac{1}{2}} + a_1 E + a_2 E^{\frac{3}{2}}) + a_3 E^{\frac{1}{2}}\delta f, \tag{6}$$

where $a_0 = \frac{2\sqrt{2}}{\pi}\cdot\frac{\Phi}{\sqrt{J}}$, $a_1 = \frac{\mu}{J} < 0$, $a_2 = \frac{8\sqrt{2}}{3\pi}\cdot\frac{\kappa}{J^{\frac{3}{2}}}$, $a_3 = \frac{2\sqrt{2}}{\pi}\cdot\frac{1}{\sqrt{J}}$.

The minus sign on the right side of equation (6) corresponds to the plus sign in formula (5). Equation (6), in which there is no fluctuating source,

$$\frac{dE}{dt} = \mp(a_0 E^{\frac{1}{2}} + a_1 E + a_2 E^{\frac{3}{2}}) \tag{7}$$

has three stationary points $E = 0$ and

$$\sqrt{E_{1,2}} = \frac{-a_1 \pm \sqrt{a_1^2 - 4a_0 a_2}}{2a_2}, \tag{8}$$

where the discriminant $a_1^2 - 4a_0 a_2 = \frac{(\mu^2 - \delta_K^2)}{J^2}$; $\delta_K^2 = \frac{128\Phi\kappa}{3\pi^2}$.

Let us show that for $\delta_K^2 < \mu^2$ the dynamic system (Keldysh model) relaxes either to an external $\sqrt{E_1}$, either to the internal $\sqrt{E_2}$ limit cycle depending on the minus or plus sign in equation (7).

The solution to equation (7), on the right-hand side of which there is a minus sign, has the form

$$\sqrt{E} = \frac{\sqrt{E_1} - \sqrt{E_2} A \cdot exp(-t/\tau)}{1 - A \cdot exp(-t/\tau)}, \qquad (9)$$

where $A = \dfrac{\sqrt{E_0} - \sqrt{E_1}}{\sqrt{E_0} - \sqrt{E_2}}$; $E_0 = E(t_0 = 0)$ – the initial value of the energy of a dynamical system (Keldysh model with one degree of freedom) at the moment of time $t_0 = 0$; the relaxation time to a stationary state has the form

$$\tau = \frac{2J}{\sqrt{\mu^2 - \delta_K^2}}. \qquad (10)$$

It is seen that at $t \gg \tau$ dynamic system (Keldysh model) relaxes to external $\sqrt{E_1}$ limit cycle. The solution for equation (7), on the right-hand side of which there is a plus sign, coincides with formula (9) with the replacement $t \to -t$. It is seen that in this case the dynamical system (the Keldysh model) relaxes to the internal $\sqrt{E_2}$ limit cycle.

## 3. Calculation results

The stationary solution of equation (6), on the right-hand side of which there is a minus sign, is constructed using (see, for example, [7])

$$g(E) = \frac{1}{N\sqrt{E}} \exp\left[-\left(\frac{2a_0 E^{\frac{1}{2}} + a_1 E + \frac{2}{3} a_2 E^{\frac{3}{2}}}{Da_3^2}\right)\right], \qquad (11)$$

where $N$ – normalization factor.

The graph of the function (11) of the probability density of detecting a dynamic system (Keldysh model) depending on the energy of the dynamic system is constructed for two values of the intensity of a random source $D = 0.2$ in Fig. 1 and $D = 0.4$ in Fig. 2. Constants $J = k = \kappa = 1; \mu = -1.2987; \Phi = 0.2$ are defined in [2].

Using formula (8), we can calculate $E_1 \approx 0.845$ and $E_2 \approx 0.027$. The radii of the outer and inner limit cycle in Fig. 1a in [2] are close to the values $\sqrt{2E_1} \approx 1.303$ and $\sqrt{2E_2} \approx 0.23$, respectively. It can be seen that the graph of function (11) in Fig. 1 reaches its maximum value at the energy of the dynamic system $E = 0$ and with an energy that is slightly less than $E_1 \approx 0.845$. A similar result was obtained in [5] when studying the effect of noise intensity on the position of the maximum of the probability density of detecting a dynamical system, which is described by the van der Pol equation.

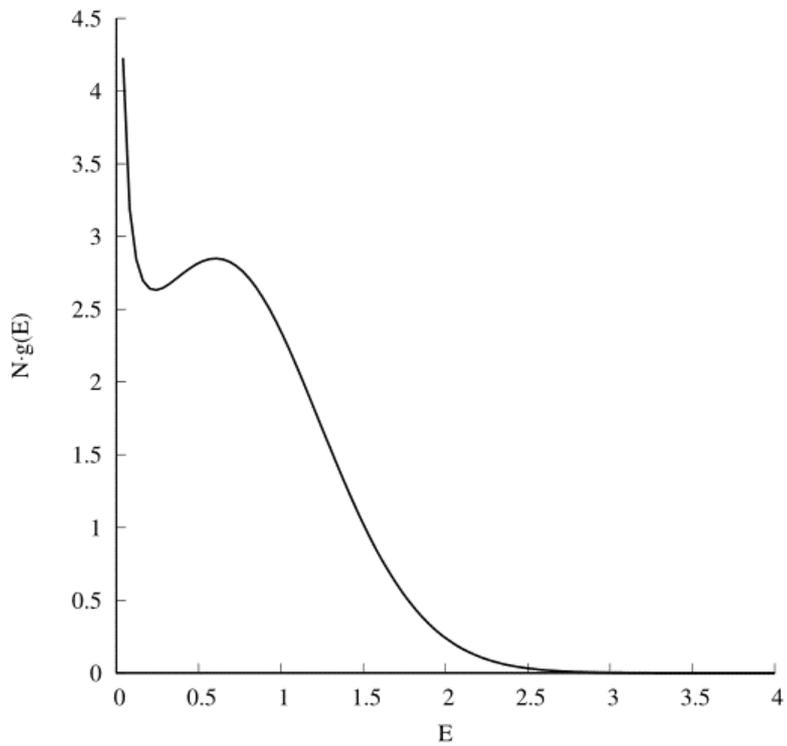

**Fig. 1.** The graph of the probability density of detecting a dynamic system (Keldysh model) depending on the energy of the dynamic system is plotted for the value of the intensity of a random source $D = 0.2$.

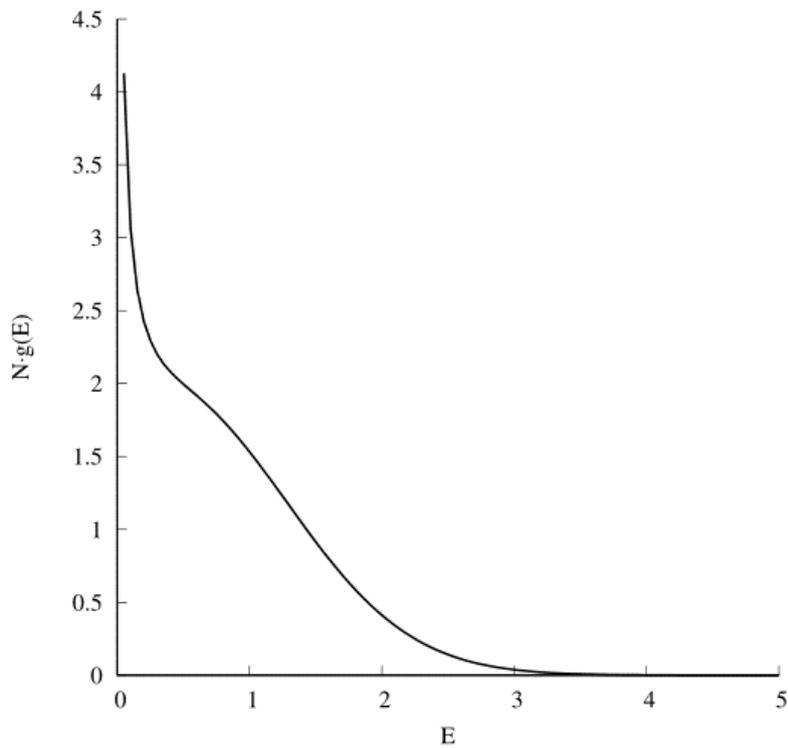

**Fig. 2.** The graph of the probability density of detecting a dynamic system (Keldysh model) depending on the energy of the dynamic system is plotted for the value of the intensity of a random source $D = 0.4$.

## 4. CONCLUSION

It is seen that at $D \geq D_{crit.} \approx 0.4$ self-oscillations are suppressed, since the maximum of function (11) disappeared, which describes the probability density of detecting a dynamical system in the vicinity of the limit cycle. The dynamical system passes into a quasi-stationary state of rest, since self-oscillations reappear with a decrease in the intensity of a random source.

## 5. ACKNOWLEDGMENTS

The reported study was funded by RFBR, project number 20-07-00236 a.